\documentclass[a4paper,11pt]{article}
\usepackage{pos}
\usepackage{here}
\usepackage{url}
\usepackage{graphicx}
\usepackage{dcolumn}
\usepackage{bm}
\usepackage{comment}

\usepackage{braket}

\title{Gravitational form factors of the nucleon from the chiral effective model}

\author*[a,b]{Daisuke Fujii}
\author[c]{Mamiya Kawaguchi}
\author[d]{Mitsuru Tanaka}

\affiliation[a]{Advanced Science Research Center, Japan Atomic Energy Agency (JAEA), \\
Tokai, 319-1195, Japan}

\affiliation[b]{Research Center for Nuclear Physics, Osaka University, \\
Ibaraki 567-0048, Japan}

\affiliation[c]{Center for Fundamental Physics, School of Mechanics and Physics, Anhui University of Science and Technology, Huainan, 232001, People’s Republic of China}

\affiliation[d]{Department of Physics, Nagoya University, Nagoya 464-8602, Japan, \\ Tokai, Ibaraki 319-1106, Japan}

\emailAdd{daisuke@rcnp.osaka-u.ac.jp}

\abstract{
We investigate the confining pressure and associated global property, i.e, D-term, 
of the nucleon using the skyrmion approach formulated within scale-invariant chiral perturbation theory. In this framework, the nucleon is modeled as a skyrmion, and a scalar meson is introduced to incorporate the effects of the scale anomaly via low-energy theorems. The contributions from the current quark mass and gluonic dynamics to the scale anomaly are encoded through the pion and scalar meson masses, respectively.
By decomposing the nucleon's energy-momentum tensor, we isolate the anomalous components and analyze their role in generating pressure. We find that the gluonic contribution to the scale anomaly plays a dominant role in producing confining pressure. In comparison with results from conventional chiral perturbation theory in the chiral limit, the total pressure derived from sChPT provides improved qualitative agreement with lattice QCD results. We also evaluate the D-term and compare it with recent lattice and model-independent determinations.}

\FullConference{The 21st International Coference on Hadron Spectroscopy and Structure (HADRON2025)\\
27-31 March, 2025\\
Toyonaka Campus, Osaka University, Japan\\}

\begin{document}
\maketitle

\section{Introduction}

Understanding how quarks and gluons are confined within hadrons remains a central challenge in quantum chromodynamics (QCD). One promising approach to this problem is to investigate the internal stress distribution—namely, pressure and shear forces—encoded in the energy-momentum tensor (EMT) and accessed through gravitational form factors (GFFs). Since the first experimental extraction of the proton’s pressure distribution in 2018~\cite{Burkert:2018bqq}, there has been growing theoretical interest in how such mechanical properties reflect non-perturbative features of QCD~\cite{Fujii:2024rqd,Ji:2025gsq,Fujii:2025aip,Fujii:2025tpk,Tanaka:2025pny,Fujii:2025paw}, including condensates and symmetry breaking. Despite such progress, the specific QCD mechanisms that govern the confining pressure and influence the GFFs remain unclear.

To shed light on this issue, we focus on the scale anomaly—a fundamental non-perturbative feature of QCD—and examine its impact on the pressure obtained from the GFFs of the nucleon. 
The scale anomaly reflects the breaking of scale invariance by explicit quark mass terms (with anomalous dimension $\gamma_m$) and gluonic quantum effects. 
As we will demonstrate, this anomaly plays a significant role in generating the negative pressure--confining pressure--in the nucleon. 

To investigate the role of the scale anomaly in hadron structure, we employ the Skyrme model 
formulated within scale-invariant chiral perturbation theory (sChPT), which incorporates the scale anomaly
through a scalar meson coupled to the dilatation current. This framework extends conventional chiral perturbation theory (ChPT) by including 
the anomalous breaking of scale invariance alongside chiral symmetry. While previous studies have analyzed the pressure by separately considering the contributions from quark and gluon dynamics, the specific role of the scale anomaly has remained less explored. In this proceedings, based on Ref.~\cite{Fujii:2025aip}, we show that the confining pressure inside the nucleon is primarily generated by the gluonic component of the scale anomaly. The resulting pressure, evaluated in sChPT, exhibits improved agreement with lattice QCD~\cite{Shanahan:2018nnv} compared to results based on conventional ChPT~\cite{Cebulla:2007ei}. 
Furthermore, we calculate the D-term in this model and obtain a result in good agreement with recent lattice QCD data.
Furthermore, we compute the D-term within the framework of this model and obtain a value consistent with recent lattice data~\cite{Hackett:2023rif} and model-independent analyses~\cite{Cao:2024zlf}.

\section{Formulation}

We start with the matrix elements of the EMT for nucleon states:
\begin{align}
    &\Braket{p',s'|\Theta_{\mu\nu}(0)|p,s}=\bar{u}'\Big[A(t)\frac{P_\mu P_\nu}{M_N}+J(t)\frac{i\left(P_\mu\sigma_{\nu\rho}+P_\nu\sigma_{\mu\rho}\right)\Delta^\rho}{2M_N}+D(t)\frac{\Delta_\mu\Delta_\nu-g_{\mu\nu}\Delta^2}{4M_N}\Big]u, \label{GFFs}
\end{align}
where $A(t), \ J(t), \ D(t)$ are the GFFs; $\sigma_{\mu\nu}$ is defined as $\sigma_{\mu\nu}=\frac{i}{2}[\gamma^\mu,\gamma^\nu]$ with the Dirac matrix $\gamma_\mu$; the momentum $P^\mu$ and $\Delta^\mu$ are defined by $P^\mu=(p^\mu+p'^\mu)/2$ and $\Delta^\mu=p'^\mu-p^\mu$, and $t=\Delta^2$;
$M_N$ is a nucleon mass; 
$s,s'=\pm1/2$ represents the spin state of a hadron;
$u(p,s)$ is the Dirac spinor. 

The spatial distribution of the nucleon’s EMT, referred to as the static EMT $\Theta^{\rm static}_{\mu\nu}(\vec{r},\vec{s})$, is investigated through its Fourier transform in the Breit frame, where $P_\mu = (E, \vec{0})$ and $\Delta_\mu = (0, \vec{\Delta})$. To isolate the contribution associated with the scale anomaly, the EMT is decomposed into a trace and a traceless component as
\begin{align}
    &\Theta^{\rm static}_{\mu\nu}(\vec{r},\vec{s})=\bar{\Theta}^{\rm static}_{\mu\nu}+\hat{\Theta}^{\rm static}_{\mu\nu} \label{staticEMTdecomp} \\
    &\bar{\Theta}^{\rm static}_{\mu\nu}(\vec{r},\vec{s})=\int\frac{d^3\vec{\Delta}}{(2\pi)^3}e^{-i\vec{r}\cdot\vec{\Delta}}\frac{\Braket{p',s'|\bar{\Theta}_{\mu\nu}(0)|p,s}}{\bar u(p')u(p)}\\
    &\hat{\Theta}^{\rm static}_{\mu\nu}(\vec{r},\vec{s})=\int\frac{d^3\vec{\Delta}}{(2\pi)^3}e^{-i\vec{r}\cdot\vec{\Delta}}\frac{\Braket{p',s'|\hat{\Theta}_{\mu\nu}(0)|p,s}}{\bar u(p')u(p)}.
\end{align}
Here, the traceless part is defined as $\bar{\Theta}_{\mu\nu} = \Theta_{\mu\nu} - \frac{1}{4}g_{\mu\nu}{\Theta^\rho}_\rho$, which corresponds to the dynamical contribution of quarks and gluons, 
whereas the trace part $\hat{\Theta}_{\mu\nu} = \frac{1}{4}g_{\mu\nu}{\Theta^\rho}_\rho$ represents the contribution from the QCD scale anomaly.
We adopt the normalization $\bar u(p')u(p) = 2E$ with $E = \sqrt{M_N^2 + \vec{\Delta}^2/4}$ and $\langle p'|p \rangle = 2p^0(2\pi)^3\delta^{(3)}(\vec{p}'-\vec{p})$, and focus on the case $s = s'$. The pressure distribution is defined as $p(r) = \delta^{ij} \Theta^{\rm static}_{ij}(r)/3$. The conservation of the static EMT, $\partial^i \Theta^{\rm static}_{ij} = 0$, then implies the von Laue condition, $\int_0^\infty dr r^2 p(r) = 0$, which ensures the mechanical stability of the system. 

The stability of the nucleon is encoded in the spatial components of its GFFs, specifically in the D-term. This quantity is directly related to the pressure distribution through the integral
\begin{align}
    D=M_N\int d^3rr^2p(r). \label{D}
\end{align}
A negative value of the D-term, as typically found in stable hadronic systems. 

\section{Static EMT of skyrmion based on sChPT}

In this study, we employ the Skyrme model
based on sChPT~\cite{Lee:2003eg}. This model extends conventional ChPT by incorporating both chiral and scale symmetry breaking. The sChPT Lagrangian is given by
\begin{align}
    &\mathcal{L}_{\rm sChPT}=\left(\frac{\chi}{f_\phi}\right)^2\frac{f_\pi^2}{4}g^{\mu\nu}{\rm Tr}\left(\partial_\mu U^\dagger\partial_\nu U\right)+\frac{1}{32e^2}{\rm Tr}([U^\dagger\partial_\mu U,U^\dagger\partial_\nu U]^2) \notag \\
    &\hspace{10mm}+\left(\frac{\chi}{f_\phi}\right)^{3-\gamma_m}\frac{1}{4}f_\pi ^2m_\pi^2{\rm Tr}\left(U+U^\dagger\right)+\frac{1}{2}g^{\mu\nu}\partial_\mu\chi\partial_\nu\chi-\frac{1}{4}m_{\phi0}^2f_\phi^2\left(\frac{\chi}{f_\phi}\right)^4\left[\ln\frac{\chi}{f_\phi}-\frac{1}{4}\right]
\end{align}
where $U=\exp(i\pi^a\tau^a/f_\pi)$ is the chiral field and $\chi=f_\phi e^{\phi/f_\phi}$ is the conformal compensator. This Lagrangian is constructed to reproduce both the partially conserved axial current (PCAC) relation, $\braket{0|\partial_\mu J_5^{a\mu}(x)|\pi^b(p)} = -f_\pi m_\pi^2 e^{-ip\cdot x} \delta^{ab}$, and the partially conserved dilatation current (PCDC) relation, $\braket{0|\partial_\mu J_D^\mu(x)|\phi(p)} = -f_\phi m_\phi^2 e^{-ip\cdot x}$. The scalar meson $\phi$ is interpreted as the lightest isoscalar state coupled to the dilatation current. 
The model also yields the vacuum expectation value of the divergence of the dilatation current as $\Braket{0|\partial^\mu J^D_\mu(x)|0} = -\left(1+\gamma_m\right)f_\pi^2m_\pi^2 - f_\phi^2 m_{\phi0}^2/4$, 
where the first term corresponds to the quark mass contribution, consistent with the Gell-Mann-Oakes-Renner relation $f_\pi^2 m_\pi^2 = -m_f\braket{0|\bar{q}q|0}$, while the second term reflects the gluonic component of the scale anomaly.

The nucleon is described as a static soliton solution known as a skyrmion, characterized by a topological charge corresponding to the baryon number. To construct the soliton, we adopt the hedgehog ansatz for the chiral field, $U(\vec{x}) = \exp(i\vec{\tau} \cdot \hat{r} F(r))$, and assume the conformal compensator takes the form $\chi(\vec{x}) = f_\phi C(r)$, with $F(r)$ and $C(r)$ being dimensionless profile functions. Imposing boundary conditions $F(0) = \pi$, $F(\infty) = 0$, $C(\infty) = \chi_0/f_\phi$, and $dC(0)/dr = 0$, where $\chi_0$ is determined from the stationary condition at infinity, we numerically solve the coupled differential equations for $F(r)$ and $C(r)$ to obtain the skyrmion with baryon number one.

To connect the divergence of the dilatation current with the trace of the EMT, ${\Theta^\mu}_\mu = \partial_\mu J_D^\mu$, 
the EMT is improved as $\Theta_{\mu\nu} = T_{\mu\nu} + \theta_{\mu\nu}$. 
The traceless part of the EMT includes kinetic terms of the pion and scalar fields, while the trace part is separated into quark and gluon contributions in this model. The total EMT is written as $\Theta_{\mu\nu} = \bar{\Theta}_{\mu\nu} + \hat{\Theta}^q_{\mu\nu} + \hat{\Theta}^g_{\mu\nu}$ with 
\begin{align}
    &\bar{\Theta}_{\mu\nu}=\frac{f_\pi^2}{2}\left(\frac{\chi}{f_\chi}\right)^2{\rm Tr}\left(\partial_\mu U^\dagger\partial_\nu U\right)+\frac{2}{3}\partial_\mu\chi\partial_\nu\chi-\frac{1}{3}\chi\partial_\mu\partial_\nu\chi \notag \\
    &\hspace{12mm}-g_{\mu\nu}\left\{\frac{f_\pi^2}{8}\left(\frac{\chi}{f_\chi}\right)^2{\rm Tr}\left(\partial_\rho U^\dagger\partial^\rho U\right)+\frac{1}{6}\partial_\rho\chi\partial^\rho\chi-\frac{1}{12}\chi\partial_\rho\partial^\rho\chi\right\}. \label{barEMTsChPT} \\
    &\hat{\Theta}_{\mu\nu}^q=-\frac{1}{4}g_{\mu\nu}
    (1+\gamma_m)
    \frac{f_\pi^2m_\pi^2}{4}\left(\frac{\chi}{f_\chi}\right)^{3-\gamma_m}{\rm Tr}\left(U+U^\dagger\right), \label{hatEMTsChPTq} \\
    &
    \hat{\Theta}_{\mu\nu}^g=-\frac{1}{4}g_{\mu\nu}\frac{m_{\phi0}^2f_\phi^2}{4}\left(\frac{\chi}{f_\phi}\right)^4.
    \label{hatEMTsChPTg}
\end{align}
Following Ref.~\cite{GarciaMartin-Caro:2023klo}, the static EMT of the nucleon is obtained by substituting the classical Skyrmion solution into the EMT. 
To ensure that the energy density and pressure vanish at spatial infinity, we subtract constant terms from the EMT components following Refs.~\cite{Fujii:2025aip,Tanaka:2025pny}.

\section{Results}

\begin{figure*}
    \includegraphics[scale=0.37]{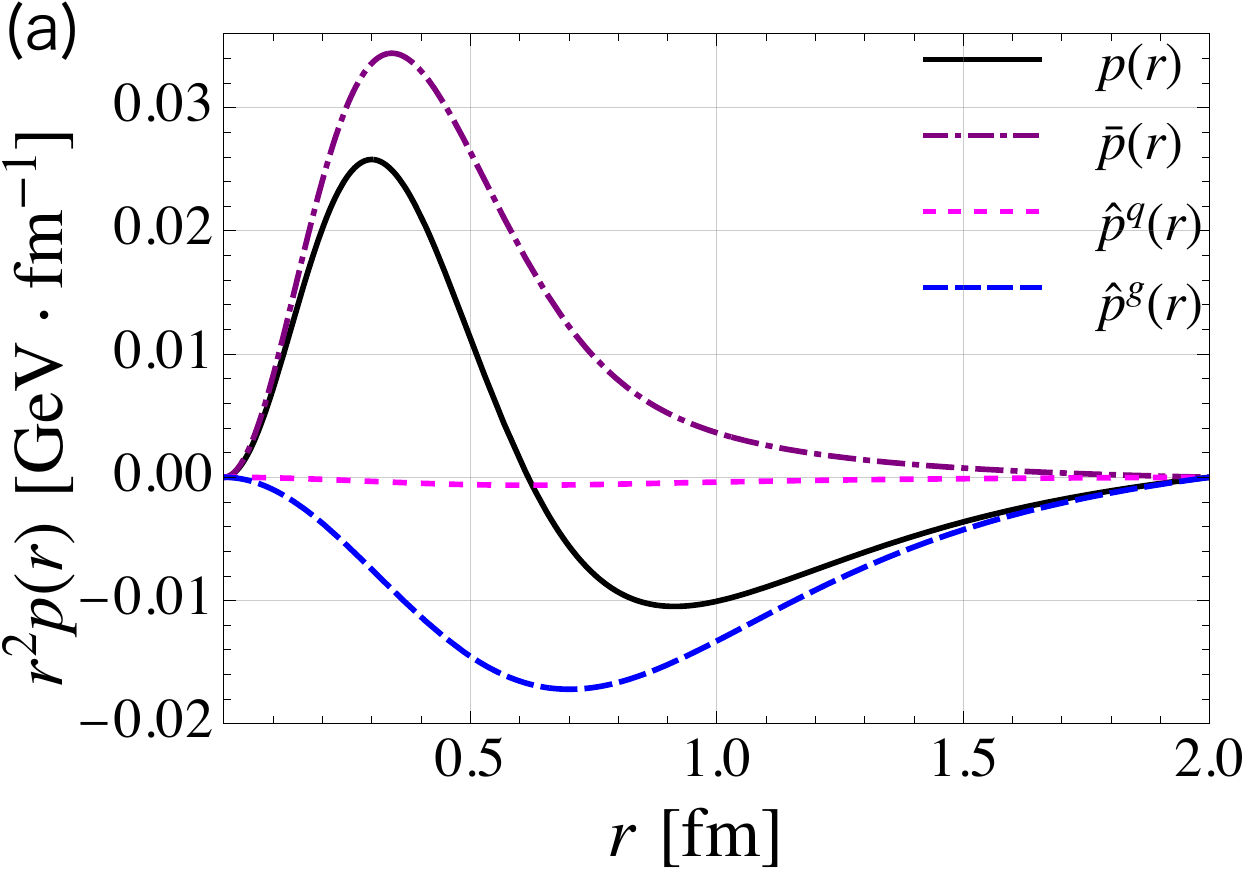}
    \includegraphics[scale=0.33]{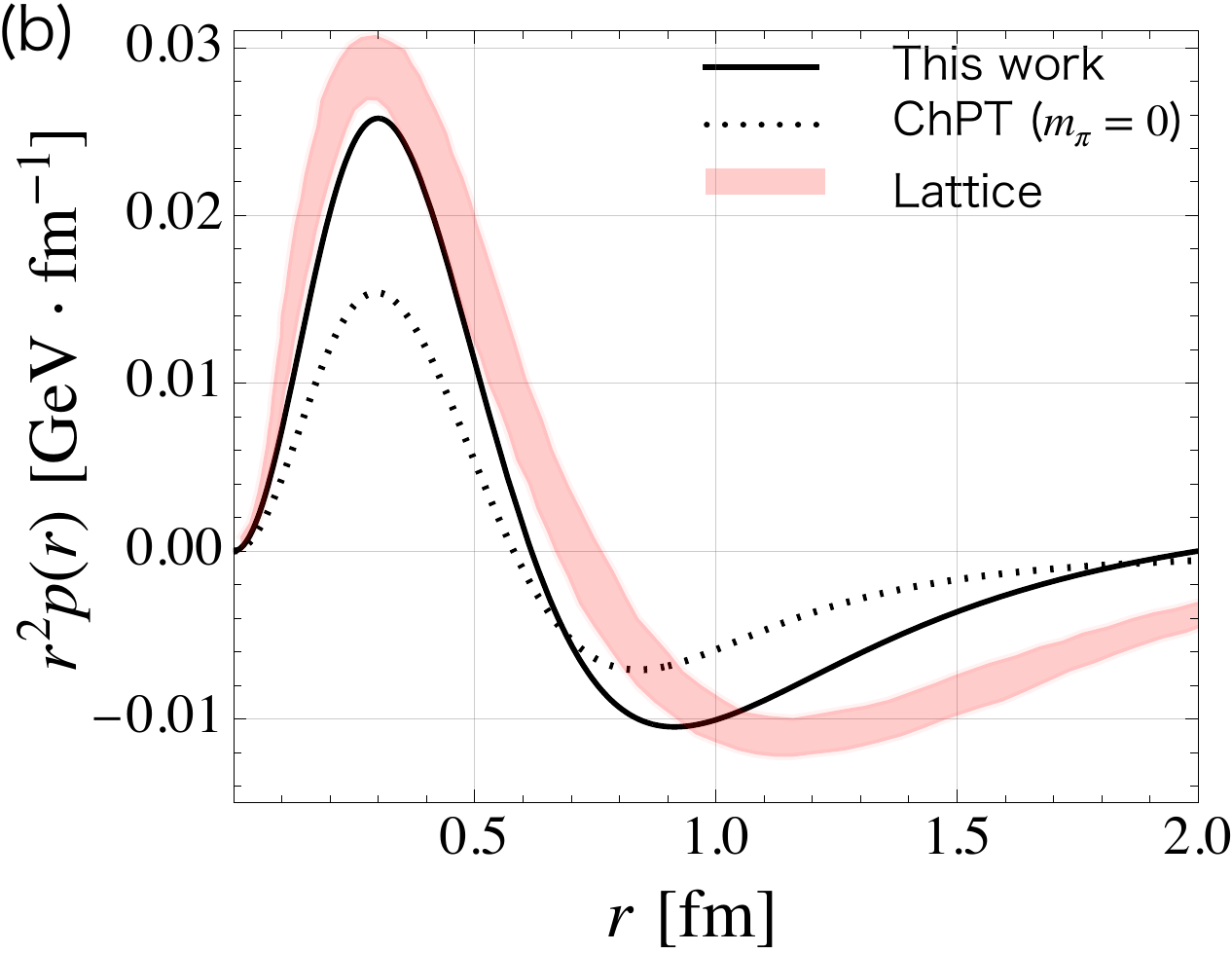}
    \caption{(a) The pressure $p(r)$ inside the nucleon and its decomposition into dynamical and anomalous parts. (b) Comparison of the total pressure in sChPT with lattice QCD data and the result from conventional ChPT. This figure is adapted from Ref.~\cite{Fujii:2025aip}.}
    \label{pressure_decomp}
\end{figure*}

We adopt the model parameters as follows. The pion decay constant and Skyrme parameter are set to $f_\pi = 68\ \text{MeV}$ and $e = 5.45$, consistent with Ref.~\cite{Adkins:1983ya}, where they are chosen to reproduce the nucleon and delta masses. The pion mass is fixed at its experimental value, $m_\pi = 140\ \text{MeV}$. For the scalar meson, we use $f_\phi = 240\ \text{MeV}$ and $m_{\phi0} = 720\ \text{MeV}$ as in Ref.~\cite{Lee:2003eg}; the decay constant is empirically suited to describe skyrmions within the scale-invariant framework, and the mass corresponds to the lightest scalar meson. The anomalous dimension of the quark mass is neglected by setting $\gamma_m = 0$, as its nonzero values do not affect the main conclusions.

Figure~\ref{pressure_decomp} (a) shows the pressure $p(r)$ inside the nucleon and its decomposition into dynamical and anomalous contributions. The dynamical part, $\bar{p}(r)$, is positive across all spatial regions and corresponds to a repulsive force. In contrast, the anomalous part, $\hat{p}^{q,g}(r)$, is negative and provides a confining force, with the gluonic contribution being dominant. Remarkably, the gluonic contribution remains substantial even at large distances, eventually outweighing the repulsive part and driving the total pressure negative as $r$ increases.

To evaluate the validity of our approach, we compare the total pressure from sChPT with lattice QCD results~\cite{Shanahan:2018nnv} in Fig.~\ref{pressure_decomp}(b). For reference, we also show the Skyrme model result based on ChPT in the chiral limit ($m_\pi = 0$), which lacks a proper account of the scale anomaly. The ChPT result significantly underestimates the pressure at short distances, whereas sChPT better reproduces the lattice data in the small-$r$ region. At larger distances, some deviations appear, possibly due to the relatively heavy pion mass used in the lattice calculation.

Additionally, we estimate the D-term from Eq.~\ref{D} as
\begin{align}
D = -4.12. \label{Dvalue}
\end{align}
This value is consistent with recent lattice QCD results within systematic uncertainties based on dipole fits~\cite{Hackett:2023rif}, although it is slightly smaller than estimates from z-expansion fits and model-independent analyses using dispersion relations~\cite{Cao:2024zlf}. One possible origin of this discrepancy is the choice of $\gamma_m = 0$ in our model. 
We have verified that introducing a nonzero $\gamma_m$ reduces the magnitude of the negative D-term, as detailed in Ref.~\cite{Tanaka:2025pny}. 

\section{Conclusion}

In this proceedings, we investigated how the scale anomaly influences the pressure distribution inside the nucleon, using the skyrmion framework based on sChPT. Our results demonstrate that the gluonic contribution to the scale anomaly plays a leading role in generating the confining pressure. While internal stress distributions of hadrons have been widely studied, a detailed analysis isolating the scale anomaly via the decomposition in Eq.~(\ref{staticEMTdecomp}) had not been carried out. To further clarify its role in the confining pressure and D-term, extended studies using lattice QCD and effective models—both for the nucleon and other hadrons—would be highly valuable.

\section*{Acknowledgments}

The author M.T. would like to take this opportunity to thank the financial support from "THERS Make New Standards Program for the Next Generation Researchers" and JST SPRING, Grant Number JPMJSP2125.
This work of D.F. was supported in part by the Japan Society for the Promotion of Science (JSPS) KAKENHI (Grants No. JP24K17054).
The work of M.K. is supported by RFIS-NSFC under Grant No. W2433019.

\bibliographystyle{JHEP}
\bibliography{reference}

\providecommand{\href}[2]{#2}\begingroup\raggedright\begin{thebibliography}{10}

\bibitem{Burkert:2018bqq}
V.D.~Burkert, L.~Elouadrhiri and F.X.~Girod, \emph{{The pressure distribution
  inside the proton}},
  \href{https://doi.org/10.1038/s41586-018-0060-z}{\emph{Nature} {\bfseries
  557} (2018) 396}.

\bibitem{Fujii:2024rqd}
D.~Fujii, A.~Iwanaka and M.~Tanaka, \emph{{Gravitational form factors of pion
  from top-down holographic QCD}},
  \href{https://doi.org/10.1103/PhysRevD.110.L091501}{\emph{Phys. Rev. D}
  {\bfseries 110} (2024) L091501}
  [\href{https://arxiv.org/abs/2407.21113}{{\ttfamily 2407.21113}}].

\bibitem{Ji:2025gsq}
X.~Ji and C.~Yang, \emph{{Momentum Flow and Forces on Quarks in the Nucleon}},
  \href{https://arxiv.org/abs/2503.01991}{{\ttfamily 2503.01991}}.

\bibitem{Fujii:2025aip}
D.~Fujii, M.~Kawaguchi and M.~Tanaka, \emph{{Dominance of gluonic scale anomaly
  in confining pressure inside nucleon and D-term}},
  \href{https://doi.org/10.1016/j.physletb.2025.139559}{\emph{Phys. Lett. B}
  {\bfseries 866} (2025) 139559}
  [\href{https://arxiv.org/abs/2503.09686}{{\ttfamily 2503.09686}}].

\bibitem{Fujii:2025tpk}
D.~Fujii, A.~Iwanaka and M.~Tanaka, \emph{{Dominance of scale anomaly in
  confining pressure inside pions on light front in the top-down holographic
  QCD}},  \href{https://arxiv.org/abs/2507.18690}{{\ttfamily 2507.18690}}.

\bibitem{Tanaka:2025pny}
M.~Tanaka, D.~Fujii and M.~Kawaguchi, \emph{{Gravitational form factors of the
  nucleon in the Skyrme model based on scale-invariant chiral perturbation
  theory}},  \href{https://arxiv.org/abs/2507.21220}{{\ttfamily 2507.21220}}.

\bibitem{Fujii:2025paw}
D.~Fujii and M.~Tanaka, \emph{{Scale-anomaly-induced confining pressure within
  hadrons}},  \href{https://arxiv.org/abs/2507.23786}{{\ttfamily 2507.23786}}.

\bibitem{Shanahan:2018nnv}
P.E.~Shanahan and W.~Detmold, \emph{{Pressure Distribution and Shear Forces
  inside the Proton}},
  \href{https://doi.org/10.1103/PhysRevLett.122.072003}{\emph{Phys. Rev. Lett.}
  {\bfseries 122} (2019) 072003}
  [\href{https://arxiv.org/abs/1810.07589}{{\ttfamily 1810.07589}}].

\bibitem{Cebulla:2007ei}
C.~Cebulla, K.~Goeke, J.~Ossmann and P.~Schweitzer, \emph{{The Nucleon
  form-factors of the energy momentum tensor in the Skyrme model}},
  \href{https://doi.org/10.1016/j.nuclphysa.2007.08.004}{\emph{Nucl. Phys. A}
  {\bfseries 794} (2007) 87}
  [\href{https://arxiv.org/abs/hep-ph/0703025}{{\ttfamily hep-ph/0703025}}].

\bibitem{Hackett:2023rif}
D.C.~Hackett, D.A.~Pefkou and P.E.~Shanahan, \emph{{Gravitational Form Factors
  of the Proton from Lattice QCD}},
  \href{https://doi.org/10.1103/PhysRevLett.132.251904}{\emph{Phys. Rev. Lett.}
  {\bfseries 132} (2024) 251904}
  [\href{https://arxiv.org/abs/2310.08484}{{\ttfamily 2310.08484}}].

\bibitem{Cao:2024zlf}
X.-H.~Cao, F.-K.~Guo, Q.-Z.~Li and D.-L.~Yao, \emph{{Dispersive Determination
  of Nucleon Gravitational Form Factors}},
  \href{https://doi.org/10.1038/s41467-025-62278-9}{\emph{Nature Commun.}
  {\bfseries 16} (2025) 6979}
  [\href{https://arxiv.org/abs/2411.13398}{{\ttfamily 2411.13398}}].

\bibitem{Lee:2003eg}
H.-J.~Lee, B.-Y.~Park, M.~Rho and V.~Vento, \emph{{Sliding vacua in dense
  skyrmion matter}},
  \href{https://doi.org/10.1016/S0375-9474(03)01626-9}{\emph{Nucl. Phys. A}
  {\bfseries 726} (2003) 69}
  [\href{https://arxiv.org/abs/hep-ph/0304066}{{\ttfamily hep-ph/0304066}}].

\bibitem{GarciaMartin-Caro:2023klo}
A.~Garcia Martin-Caro, M.~Huidobro and Y.~Hatta, \emph{{Gravitational form
  factors of nuclei in the Skyrme model}},
  \href{https://doi.org/10.1103/PhysRevD.108.034014}{\emph{Phys. Rev. D}
  {\bfseries 108} (2023) 034014}
  [\href{https://arxiv.org/abs/2304.05994}{{\ttfamily 2304.05994}}].

\bibitem{Adkins:1983ya}
G.S.~Adkins, C.R.~Nappi and E.~Witten, \emph{{Static Properties of Nucleons in
  the Skyrme Model}},
  \href{https://doi.org/10.1016/0550-3213(83)90559-X}{\emph{Nucl. Phys. B}
  {\bfseries 228} (1983) 552}.

\end{thebibliography}\endgroup



\end{document}